\documentclass[11pt]{article}
\usepackage{epsfig}
\begin{document}  
\thispagestyle{empty}
\begin{flushright}
hep-lat/0203013
\end{flushright} 
\vskip15mm
\begin{center}
{\huge
The topological susceptibility of 
\vskip2mm
SU(3) gauge theory near $T_c$}
\vskip11mm
{\bf Christof Gattringer$^\dagger$, Roland Hoffmann and Stefan Schaefer}
\vskip3mm
Institut f\"ur Theoretische Physik, Universit\"at
Regensburg \\
D-93040 Regensburg, Germany 
\vskip20mm
\begin{abstract}  
We compute the topological susceptibility 
$\chi_t$ in SU(3) lattice gauge theory
using fermionic methods based on the Atiyah-Singer index theorem.
Near the phase transition we find a smooth crossover 
behavior for $\chi_t$ with values decreasing from (191(5) MeV)$^4$ 
to (100(5) MeV)$^4$ as we increase the temperature from 0.88 $T_c$ to 1.31 
$T_c$, showing that topological excitations exist far above $T_c$. 
Our study is the first large scale analysis of the  
topological susceptibility at high temperature
based on the index theorem and the results 
agree well with field theoretical methods.
\end{abstract}
\vskip5mm
{\sl To appear in Physics Letters B} 
\end{center}
\vfill \nopagebreak \begin{flushleft} \rule{2 in}{0.03cm}
\\ {\footnotesize \ 
${}^\dagger$ Supported by the Austrian Academy of Sciences (APART 654).}
\end{flushleft}
\newpage
\setcounter{page}{1}
Relating topology to the dynamics of a gauge theory is a fascinating  
idea which can provide deep insights to the non-perturbative structure of
the theory. For QCD an object of particular interest is the
topological susceptibility $\chi_t$ defined as 
\begin{equation}
\chi_t \; \; = \; \; \frac{1}{V} \; \langle Q^2 \rangle \; .
\label{topsusc} 
\end{equation}
$Q$ denotes the topological charge obtained
e.g.~as an integral over $F_{\mu \nu} \widetilde{F}_{\mu \nu}$. 
For QCD $\chi_t$ plays an important role in the low energy behavior where
e.g.~the so-called Witten-Veneziano formula \cite{WiVe}
relates the quenched $\chi_t$ to the
meson masses. For pure gauge theory $\chi_t$ is a measure
for the net abundance of topological fluctuations and can provide insight
to instanton liquid models \cite{instrev}. 
Here the behavior around the phase 
transition is of particular interest, since new machines such as RHIC
are starting to probe hot QCD. $\chi_t$ is expected to show
a smooth crossover behavior into the high temperature phase as 
topological excitations die out.

The computation of $\chi_t$ near $T_c$ is an inherently non-perturbative
calculation and can e.g.~be attacked in lattice gauge theory. While
for SU(3) lattice gauge theory at zero temperature there are several 
recent calculations \cite{chidet}-\cite{Aletal}, 
only a single modern study \cite{Aletal}
of the critical behavior near $T_c$ can be found in the 
literature (for an early attempt see \cite{early}). 
In this article we improve on this situation and present a systematic
study of $\chi_t$ near the critical temperature with high statistics and
an assessment of finite size effects from a comparison of data
from different lattice sizes.  

The traditional approach \cite{chidet}
to the calculation of the topological charge was a 
discretization of $F_{\mu \nu} \widetilde{F}_{\mu \nu}$ (the so-called
field theoretical method).
Lattice discretizations of $F_{\mu \nu} \widetilde{F}_{\mu \nu}$ are
very sensitive to ultraviolet fluctuations and can be used only
after some cooling or smearing procedure has been applied to the gauge field.
Such a cooling step, however, has the potential to destroy topological
lumps and to alter the outcome for $\chi_t$. Thus it is important to
check whether there is consistency with alternative methods.

In this letter we compute the topological charge using a fermionic method
(compare also \cite{chiferm,adelaide}).
For the continuum Dirac operator $D$ the Atiyah-Singer index theorem 
\cite{AtSi71} relates the
topological charge $Q$ to the index of $D$ which can be written as the 
difference of the numbers of left-handed and right-handed zero-modes. 
For the standard Wilson-Dirac operator such a determination of $Q$ through 
the zero-modes fails for numerically accessible values of the cutoff 
since the would-be zero-modes mix with the corresponding modes in the 
doubler branches \cite{wilsonindex}. 

With the re-discovery of the Ginsparg-Wilson relation \cite{GiWi82} for
lattice Dirac operators it was understood how to formulate chirally symmetric 
fermions on the lattice. In particular Ginsparg-Wilson fermions give
rise to an index theorem \cite{giwiindex} already at finite cutoff. 
It was found \cite{giwiindex,giwiindex2} that 
\begin{equation}
Q \; \; = \; \; n_- \; - \; n_+ \; \; = \; \; 
\frac{1}{2} \, \mbox{Tr} \, \gamma_5 \, D \; \; = \; \; 
\frac{1}{32\pi^2}\int d^4x F_{\mu \nu}(x) \widetilde{F}_{\mu \nu}(x) \; 
+ \; {\cal O} (a^2) \;. 
\label{lattind}
\end{equation}
In the first line of the equation $n_+$ ($n_-$) 
denotes the number of zero-modes with positive
(negative) chirality. The problem with the doublers is resolved since
their would-be zero-eigenvalues are shifted to $2/a$. The great disadvantage
of an exact solution of the Ginsparg-Wilson equation such as the overlap
operator \cite{overlap} is the high cost of a numerical implementation.

A possible, considerably cheaper approach is to use an approximate solution
of the Ginsparg Wilson equation such as a finite parameterization of the
fixed point action \cite{perfect1}. In this study 
we use chirally improved 
fermions \cite{chirimp} which arise from a systematic expansion of
a solution of the Ginsparg-Wilson equation. They are sufficiently chiral to 
allow for an unambiguous identification of the zero-modes but are 
numerically considerably cheaper than overlap fermions 
and allow to obtain the statistics necessary
for a precise measurement of $\chi_t$.

Our (quenched) SU(3) 
gauge configurations were generated with the L\"uscher-Weisz
action \cite{LuWeact}. The leading coupling $\beta_1$ varied in a range
from $\beta_1 = 8.10$ to $\beta_1 = 8.60$. When using the Sommer parameter
$r_0 = 0.5$ fm to set the scale this corresponds 
\cite{scale} to a range of
lattice spacings from $a = 0.084$ fm to $a = 0.125$ fm (compare 
Table~\ref{rundat} below). For details concerning the Monte Carlo see
\cite{Gaetal01a}.
We used two settings of lattices: Firstly, $L^4$ lattices with 
$L = 16$ and $L=12$. These ensembles are all in the low temperature phase
of QCD, where quarks are confined and chiral symmetry is broken. Secondly,
we use lattices of size $6\times L^3$ with $L=20, L=16$ and $L=12$. These 
latter ensembles give rise to temperatures ranging from values 
below the critical temperature (0.88 $T_c$) to values above the phase 
transition (1.31 $T_c$). The details for the parameters of our runs can be 
found in Table~\ref{rundat}. For the L\"uscher-Weisz action the critical 
temperature was computed in \cite{gappaper} using Polyakov loops and also 
the gap of the spectrum of the Dirac operator. Both the chiral transition
and the deconfinement transition were found to have a critical temperature
of $T_c = 300(3)$ MeV. 
\begin{table}[p]
\begin{center}
\begin{tabular}{cccccc}
\small
\\[-2mm]
lattice & $\beta_1$ & $a$ [fm] & $T$ [MeV] & $N_{confs}$ & 
$\chi_t^\frac{1}{4}$ 
[MeV] \\[2mm]
\hline \\[-2mm]
$6\times20^3$ & 8.10 & 0.125(1) & 264(2) & 400 & 194(5) \\
$6\times20^3$ & 8.20 & 0.115(1) & 287(3) & 400 & 185(5) \\
$6\times20^3$ & 8.25 & 0.110(1) & 299(3) & 400 & 162(5) \\
$6\times20^3$ & 8.30 & 0.106(1) & 311(3) & 400 & 144(5) \\
$6\times20^3$ & 8.45 & 0.094(1) & 350(4) & 400 & 124(5) \\
$6\times20^3$ & 8.60 & 0.084(1) & 392(5) & 400 &  93(5) \\[2mm]
\hline \\[-2mm]
$6\times16^3$ & 8.10 & 0.125(1) & 264(2) & 800 & 200(4) \\
$6\times16^3$ & 8.20 & 0.115(1) & 287(3) & 800 & 186(4) \\
$6\times16^3$ & 8.25 & 0.110(1) & 299(3) & 800 & 170(4) \\
$6\times16^3$ & 8.30 & 0.106(1) & 311(3) & 800 & 153(4) \\
$6\times16^3$ & 8.45 & 0.094(1) & 350(4) & 800 & 123(4) \\
$6\times16^3$ & 8.60 & 0.084(1) & 392(5) & 800 & 105(5) \\[2mm]
\hline \\[-2mm]
$6\times12^3$ & 8.10 & 0.125(1) & 264(2) & 1200 & 191(4) \\
$6\times12^3$ & 8.20 & 0.115(1) & 287(3) & 1200 & 183(4) \\
$6\times12^3$ & 8.25 & 0.110(1) & 299(3) & 1200 & 170(4) \\
$6\times12^3$ & 8.30 & 0.106(1) & 311(3) & 1200 & 153(4) \\
$6\times12^3$ & 8.45 & 0.094(1) & 350(4) & 1200 & 118(4) \\
$6\times12^3$ & 8.60 & 0.084(1) & 392(5) & 1200 & 101(5) \\[2mm]
\hline \\[-2mm]
$16^4$ & 8.10 & 0.125(1) &  99(1) & 200 & 185(6) \\
$16^4$ & 8.20 & 0.115(1) & 107(1) & 200 & 194(7) \\
$16^4$ & 8.30 & 0.106(1) & 117(1) & 200 & 189(6) \\
$16^4$ & 8.45 & 0.094(1) & 131(1) & 200 & 194(6) \\
$16^4$ & 8.60 & 0.084(1) & 147(2) & 200 & 192(7) \\[2mm]
\hline \\[-2mm]
$12^4$ & 8.10 & 0.125(1) & 132(1) & 400 & 200(5) \\
$12^4$ & 8.30 & 0.106(1) & 155(1) & 400 & 188(5) \\
$12^4$ & 8.45 & 0.094(1) & 175(2) & 400 & 191(5) \\[-2mm]
\normalsize
\end{tabular}
\end{center}
\caption{Parameters and results. We give the size of our lattices, the
inverse gauge coupling $\beta_1$, the corresponding lattice spacing $a$ in
fermi, the temperature in MeV, the number $N_{confs}$ of
configurations in our ensembles and our results 
for the topological susceptibility.
\label{rundat}}
\end{table}

\begin{figure}[t]
\vspace*{4mm}
\centerline{\epsfig{file=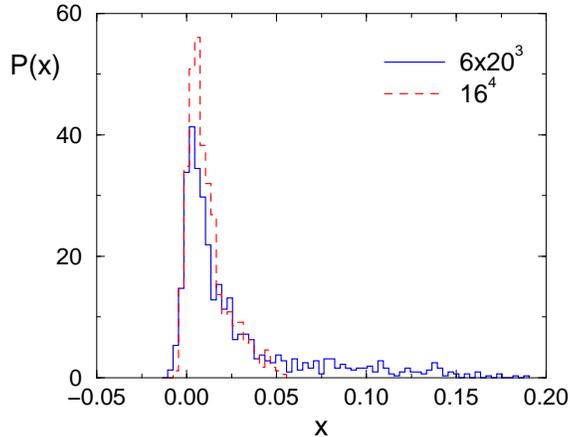,width=7.5cm}} 
\caption{ Distribution of the real eigenmodes of the chirally improved Dirac
operator for the $\beta = 8.20$ ensembles. The dashed histogram represents 
the data from the $16^4$ lattice while the full line is for $6\times20^3$.  
\label{realhist}}
\end{figure}

Since the chirally improved fermions we are using here are only an
approximation of a solution of the Ginsparg-Wilson equation a few comments
on our determination of the topological charge are in order here. 
One of the effects of using an approximation of a Ginsparg-Wilson operator
are fluctuations of the zero-eigenvalues. Instead of being located exactly 
at the origin the zero-eigenvalues manifest themselves
as small real eigenvalues. It can 
be demonstrated \cite{Gaetal01} that for discretized instantons the
size of these would-be zero-eigenvalues 
increases as the radius of the instanton 
decreases. Similarly, also for the thermalized configurations one finds 
a distribution of real eigenvalues. For two of our ensembles
($16^4$ and $6\times20^3$ both at $\beta_1 = 8.20$) we show this distribution 
in Fig.~\ref{realhist}. The distribution of the position $x$ of the real 
eigenvalues has a pronounced peak at the origin and a tail which extends 
towards larger values. The tail comes from very small instantons, so-called
dislocations.  

We compute our eigenvalues using the implicitly restarted Arnoldi method
\cite{arnoldi}. This algorithm computes a given number $N$ of eigenvalues 
ordered with respect to their absolute value. We always compute the $N=50$ 
smallest eigenvalues, except for the $16^4$ lattices where we have $N=30$. 
A potential source of error is a too low value of $N$. Sufficiently many
eigenvalues have to be computed in order to capture all of the peak
in the distribution of Fig.~\ref{realhist}. 
As is obvious from Fig.~\ref{realhist} with our choice of $N$ we obtain 
all of the peak and a large portion of the tail. 
Setting a cut for the eigenvalues at e.g.~0.1 leaves the results for 
$\chi_t$
unchanged showing that dislocations make up only a small contribution to 
the topological susceptibility. We remark that there is no danger of
mistaking a would-be zero eigenvalue for an eigenvalue with a small 
but non-vanishing imaginary part. A few lines of algebra \cite{itohetal}
show that for a $\gamma_5$-hermitian Dirac operator the matrix
element $\psi^\dagger \gamma_5 \psi$ of an eigenvector $\psi$ is
non-zero only for eigenvectors with real eigenvalues. For our approximation 
of a Ginsparg-Wilson Dirac operator we find values of 
$|\psi^\dagger \gamma_5 \psi|$ ranging from 0.8-0.9 in the peak of the
distribution down to 0.4 for the very end of the tail.

We summarize our procedure for determining $Q$ as follows: 
We compute the 50 (30 for $16^4$) eigenvalues 
of the Dirac operator and for the eigenvectors with real eigenvalues
evaluate $\psi^\dagger \gamma_5 \psi$. We take $n_+$ ($n_-$) to be the 
number of modes where $\psi^\dagger \gamma_5 \psi$ is positive (negative).
The topological charge is then computed as in Eq.~(\ref{lattind})
by $Q = n_- - n_+$.

For a subsample of 200 configurations on the $12^4$ lattices at 
$\beta_1 = 8.10$ and $\beta_1 = 8.45$ we cross-checked our procedure with
the results obtained from the overlap operator. For the overlap
operator the zero-modes have eigenvalues exactly at zero since also modes 
from dislocations are projected onto exact zero-modes. The corresponding
$\gamma_5$ matrix elements are always $\psi^\dagger \gamma_5 \psi = \pm 1$. 
When comparing the determination of $Q$ for the individual configurations
we found a discrepancy for 2\% for the configurations at $\beta_1 = 8.45$
and 9\% for $\beta_1 = 8.10$. This demonstrates that as one goes closer to 
the continuum limit the two definitions agree better. Furthermore, the 
difference in $Q$ was always only one unit such that the results for
$\chi_t$ agree surprisingly well: For the subsamples of 200 configurations
we found at $\beta_1 = 8.45$ a 
value of $\chi_t =$ (182 (7) MeV)$^4$ for both methods.
At $\beta_1 = 8.10$ we obtained a value of $\chi_t =$ (196 (7) MeV)$^4$ 
with the overlap operator and $\chi_t =$ (197 (7) MeV)$^4$ with the
chirally improved operator. Our test shows that the two methods agree
very well, however, using the overlap operator is by a factor of 10 
more costly than working with the chirally improved operator. We remark that
a direct comparison of the field theoretical
approach and the fermionic method on single
configurations can be found in \cite{adelaide}.
\begin{figure}[t]
\hspace*{-4mm}
\centerline{\epsfig{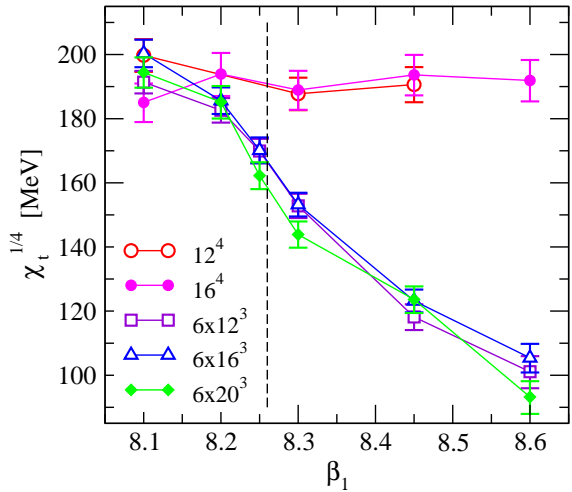}} 
\caption{ 
The topological susceptibility as a function of $\beta$. We show our results
for the zero temperature ensembles on $12^4$ (circles) and $16^4$ 
(filled circles) as well as for the finite temperature ensembles on
$6 \times 12^3, 6 \times 16^3$ and $6 \times 20^3$ lattices (squares,
triangles, diamonds). The dashed vertical line marks the phase transition.
\label{chivsbeta}}
\end{figure}

We begin with a discussion of the results for $\chi_t$ as a function of
$\beta_1$. We plot the corresponding data in Fig.~\ref{chivsbeta}.  
For time extent 6 the critical value of $\beta_1$ was determined 
\cite{gappaper} to be $\beta_1 = 8.26(2)$. We mark the transition with
a dashed vertical line in Fig.~\ref{chivsbeta}.
For the $6 \times L^3$ configurations we used values of $\beta_1$ ranging 
from $8.10$ to $8.60$ giving rise to ensembles on both sides of the 
transition. The corresponding data are represented in Fig.~\ref{chivsbeta}
by squares ($6 \times 12^3$), triangles ($6 \times 16^3$) and
diamonds ($6 \times 20^3$). We connect the symbols to guide the eye. 
When comparing the results for different volumes we find that the data
agree well within error bars. This shows that for the chosen lattice sizes 
we do not encounter finite size problems. For $\beta_1 = 8.10$
our values for $\chi_t$ scatter around (190 MeV)$^4$, which is the value
in the low temperature phase. As one increases $\beta_1$,  
$\chi_t$ starts to drop already before the critical value
$\beta_1 = 8.26$ is reached. The slope of the curve is largest near the
critical $\beta_1$. The decrease slows down as $\beta_1$ is increased 
further and $\chi_t$ 
reaches a value of $\chi_t \sim$ (100 MeV)$^4$ at $\beta_1 = 8.60$.

The decreasing curve for $\chi_t$ from the ensembles at high 
temperature can now be compared to the low temperature data 
with the same values of $\beta_1$, i.e.~the same cutoff. The corresponding 
data are represented by open circles ($12^4$) and filled circles 
($16^4$). Again we find that the data from the two different volumes
are well consistent within error bars and we do not face finite size 
problems. The numbers for $\chi_t$ remain near (190 MeV)$^4$ for 
all values of $\beta_1$. 

It is interesting to combine the
results from the zero temperature ensembles and compare them to the results 
available in the literature. The combined result from our $16^4$ lattices is
$\chi_t =$ (191(5) MeV)$^4$. The more recent results in the literature 
range between 
$\chi_t =$ (175(5) MeV)$^4$ to $\chi_t =$ (203(5) MeV)$^4$. This 
demonstrates that our zero temperature
result for the L\"uscher-Weisz action determined from
the index theorem agrees well with published data. 

\begin{figure}[t]
\hspace*{-4mm}
\centerline{\epsfig{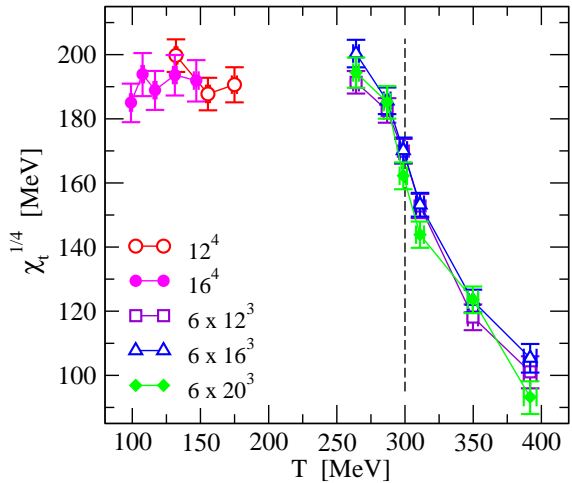}} 
\caption{The topological susceptibility as a function of the temperature.
We combine our results from all lattices: 
$12^4, 16^4, 6 \times 12^3, 6 \times 16^3$ and $6 \times 20^3$ 
(circles, filled circles, squares, triangles, diamonds). The dashed 
vertical line marks the phase transition.
\label{chivstemp}}
\end{figure}

Let us now return to the more interesting behavior near the phase transition. 
In Fig.~\ref{chivstemp} we 
present our data as a function of the temperature.
We include also our data for the $L^4$ lattices in order to have the 
baseline in the low temperature phase. 
The data from our high temperature, $6 \times L^3$ configurations
cover a temperature range from 264 MeV to 392 MeV. Again we mark the phase
transition by a vertical dashed line. As already mentioned above, the data 
from different volumes agree very well and we do not encounter finite size 
problems. The topological susceptibility starts to deviate from its zero
temperature value ($\chi_t \sim$ (190 MeV)$^4$) at a temperature of 
$T \sim 275$ MeV which is still below the deconfinement transition.  
The drop continues and becomes steepset at $T_c$ and then leans back slightly. 
At our largest temperature $T =$ 392 MeV (= 1.31 $T_c$) 
the decrease is still substantial 
and $\chi_t$ has reached $\chi_t \sim$ (100 MeV)$^4$ which is
about half of its zero temperature value. A linear extrapolation of 
the decrease of $\chi_t^{1/4}$ gives a temperature of about 600 MeV where 
$\chi_t$ becomes compatible to zero. 
This would be about twice the critical temperature, but since the
decrease of $\chi_t^{1/4}$ is certainly not linear throughout, $\chi_t \sim 0$
is presumably reached at a temperature even higher than 600 MeV. Our 
data are in reasonably well agreement with the other result for
$\chi_t$ near $T_c$ available in the literature \cite{Aletal}.

In this letter we have performed the first large scale study of
the finite temperature behavior of $\chi_t$ based on the index theorem.
Fermionic methods provide an alternative approach free from possible
ambiguities due to cooling or smearing. 
From a comparison of data on different volumes
we find that finite size effects are under good control. We find that the
topological susceptibility extends into the high temperature phase and some
topological excitations can be found up to 2 $T_c$.  
\\
\\
{\bf Acknowldegements:} 
We would like to thank Stefan D\"urr, Meinulf G\"ockeler, 
Christian Lang, Paul Rakow and Andreas Sch\"afer
for discussions. The calculations were done on the Hitachi SR8000
at the Leibniz Rechenzentrum in Munich and we thank the LRZ staff for
training and support.

\end{document}